\documentclass[manuscript]{aastex}

\usepackage{graphicx}
\usepackage{subfigure}
\usepackage{color}
\begin{document}


\title{First solar models with OPAS opacity tables }

\author{ M. Le Pennec\altaffilmark{1}, S. Turck-Chi\`eze\altaffilmark{1}, S. Salmon\altaffilmark{1},  \\ and\\ C. Blancard\altaffilmark{2}, P. Coss\' e\altaffilmark{2}, G. Faussurier\altaffilmark{2}, G. Mondet\altaffilmark{2} }

\altaffiltext{1}{CEA/IRFU/Service d'Astrophysique, CE Saclay, 91191 Gif sur Yvette, France} 
\altaffiltext{2}{CEA, DAM, DIF, F-91297 Arpajon, France}
\begin{abstract}
Stellar seismology appears more and more as a powerful tool for a better determination of the fundamental properties of solar-type stars. However the particular case of Sun is still challenging. The helioseismic sound speed determination continues to disagree with the Standard Solar Model (SSM) prediction for about a decade, questioning the reliability of this model. One of the sources of uncertainty could be in the treatment of the transport of radiation from the solar core to the surface. In this letter, we use the new
OPAS opacity tables, recently available for solar modelling, to address this issue.
We discuss first the peculiarities of these tables, then we quantify their impact on the solar sound speed and density profiles using the reduced OPAS tables  taken on the grids of the OPAL ones. We use the two evolution codes MESA and  CLES  that led to similar conclusions in the solar radiative zone. In comparison to commonly used OPAL opacity tables, the new solar models computed, for the most recent photospheric composition, with OPAS tables present improvements in the location of the base of the convective zone and in the description of the solar radiative zone in comparison to the helioseismic observations, even if the differences in the Rosseland mean opacity do not exceed 6\%. We finally carry out a comparison to a solar model computed with the OP opacity tables.
\end{abstract}


\maketitle

\section{The solar radiative zone in question}
The space missions ESA's CoRoT \citep{Baglin2006} and NASA's {\it Kepler} \citep{Gilliland2010} have already provided thousands of seismic observations of solar-like stars. This new investigation improves the knowledge of  their fundamental properties (mass, radius) with help of scaling relations \citep[and references therein]{2013_Chaplin-Miglio}. The next effort concentrates on getting an insight of their interior with help of asteroseismology. However, most of the stellar evolution codes use the same physics inputs.
It is thus important to assess the validity of these inputs to get the best scientific return of such space missions.

The Sun is a necessary test case for that purpose. The solar revised CNO  photospheric composition \citep{Asplund2005} revealed that the solar sound speed, predicted by a SSM is significantly different from the one obtained seismically from the SOHO satellite or from ground networks in the radiative zone. The differences appeared largely greater than the seismic error bars deduced with the space GOLF+MDI instruments  \citep{Turck2001,Turck2004}. 
Then, the detailed composition of the Sun has been reexamined by different groups \citep{Caffau2008,Asplund2009} but the discrepancy between the two sound speed profiles continues to be puzzling \citep{TurckPiau,Basu2015}. It reaches nearly 1\% on the sound speed, which is determined with a precision of 10$^{-4}$, that seems difficult to attribute only to the dynamical processes (direct effect of rotation or magnetic field) which are often not included in the equations describing theoretical models.

  Several hypotheses have been suggested and some of them have been quantified:
\begin{itemize}
  \item an incorrect understanding of the inner composition in part due to some elements badly known  and other part to an insufficient treatment of the microscopic diffusion, \citep{Basu2008,Basu2015}.
  \item an insufficient knowledge of the energetic balance. An upper limit of 5 \% for the possible energy difference between the energy produced by the nuclear reaction rates and the release of energy at the surface of the Sun has been estimated ~\citep[see][table 3]{TurckLopes}, in using both neutrino and seismology. If a difference exists, it could be attributed to some dynamical components not present in the energy equation of stellar structure \citep{Turck2015b}.
This idea could be checked with a very precise measurement of the pp or pep neutrino flux.
  \end{itemize}
  In this letter, we explore another hypothesis, stating that the current 
  description of the energy transport by photons is not sufficiently accurate for the interpretation of the
helio- and asteroseismic observations. If it is the case, both the use of the Rosseland mean opacity values in stellar equations
and the treatment of the microscopic diffusion in the radiative zone would be affected. 

The available opacity tables, OPAL \citep{Iglesias1996} and OP \citep{Seaton2004}, have been provided more than 10 years ago. We explore in this letter how new opacity calculations performed with the present computer resources modify the solar internal thermodynamical quantities. This first estimate uses the new OPAS tables recently available \citep{Blancard2012,Mondet2015}.

\section{The OPAS calculations}

A new generation of opacity codes is currently under development to improve the interpretation of stellar observations in the field of helio- and asteroseismology. One can  mention the ATOMIC calculations performed at Los Alamos \citep{Colgan2013}, the SCO-RCG ones performed by a CEA team \citep{Porcherot} and the OPAS ones performed by another CEA team \citep{Blancard2012}. Some outputs of these codes have been compared to a new
generation of opacity experiments performed at LULI2000 \citep{TurckLoisel2011,Turck2013,Turck2015} and on the Z machine of La Sandia \citep{Bailey2015}.

 The OPAS code  is dedicated to radiative opacity calculations of plasmas in local thermodynamic equilibrium. It is based on a detailed configuration approach \citep{Blancard2012}. The monochromatic opacity is evaluated as the sum of four different contributions involving the diffusion process,  free-free, bound-free, and bound-bound absorption processes. The bound-bound opacity is calculated by combining different approximations to take into account the level structure of configurations. Statistical or detailed methods are used to describe the transitions connecting a couple of configurations. The detailed method is based on an extensive line accounting performed in the full intermediate coupling. The bound-free opacity is evaluated using configuration-average distorted wave calculations. The free-free opacity is obtained by interpolating between the Drude-like opacity and the opacity derived from the Kramers formula including a Gaunt factor and an electron degeneracy effect correction to improve the accuracy of opacities into the complex regime where plasma and many-body effects can be important. Photon scattering by free electrons includes some collective effects as well as relativistic corrections. The different approximations and their impact on the Rosseland mean value tables are discussed \citep[see][for details]{Mondet2015} and the tables are available through this reference.
 
 \section{The OPAS tables description}
The OPAS opacity calculations are tabulated in log$_{10}$ T and log$_{10}$ R, like the OPAL tables, where 
$\rm log_{10} \,R = log_{10} \rho -3 *log_{10} \, T +18$.
For indication, the OPAL tables cover log$_{10}$ R from -8 to 1 with steps of 0.5 and log$_{10}$ T from 3.75 to 6 with steps of 0.05, from 6 to 8.1 with steps of 0.1 and from 8.1 to 8.7 by steps of 0.2. 

\begin{figure}
 \begin{center}
  \includegraphics[scale=0.30]{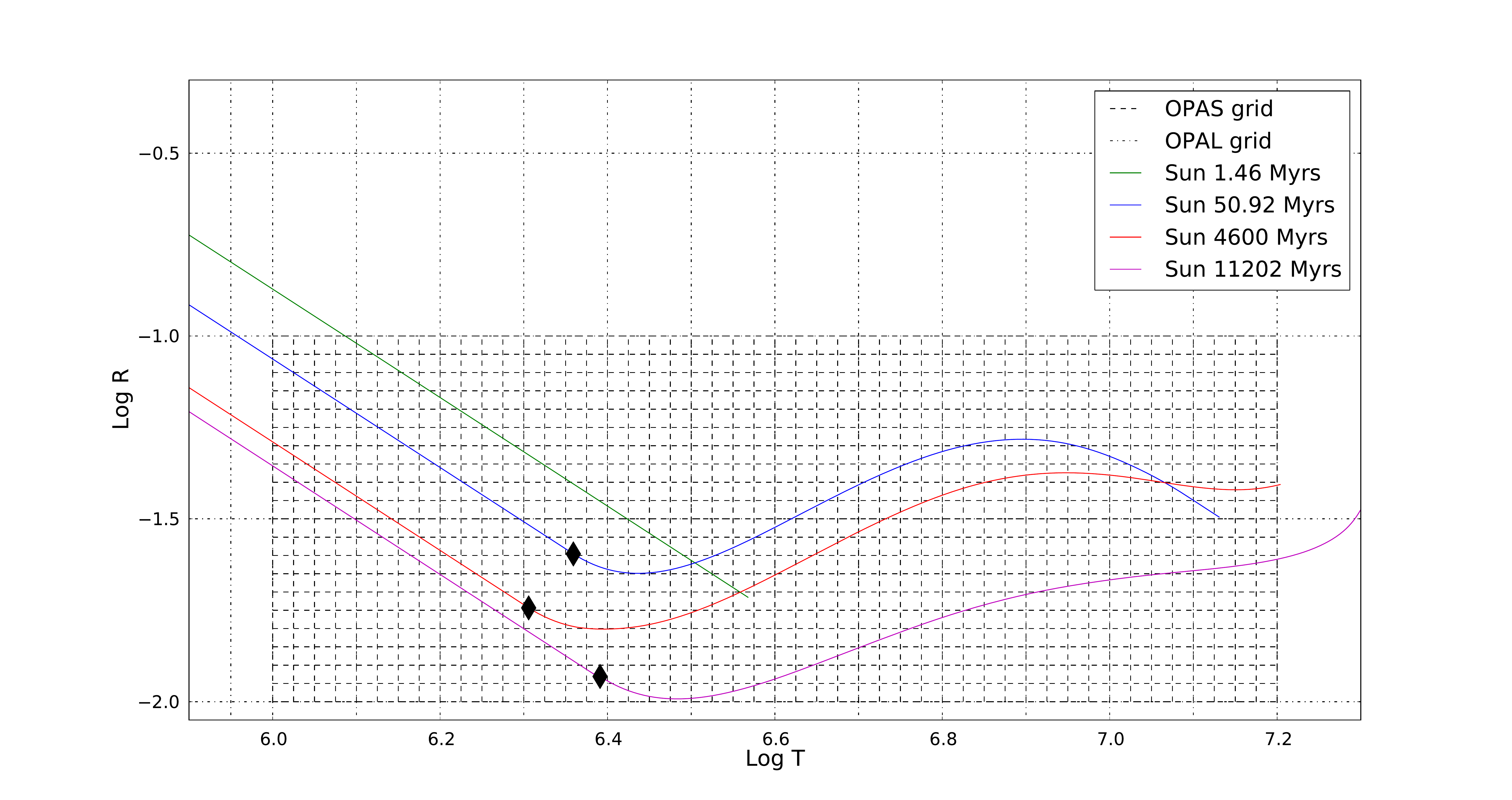} 
 \caption{OPAL and OPAS opacity meshes, superimposed to the solar path at different ages (continuous line). The diamond symbol marks the transition between the radiative and the convective zone.}
   \label{fig : new_grid}
   \end{center}
\end{figure}

The new OPAS tables are specifically dedicated to the study of the Sun  and  solar-like stars. So they have been computed with thinner grids both on log$_{10}$ T, log$_{10}$ R and Z. Consequently, for resources reasons, they are presently reduced to log$_{10}$ T from 6 to 7.2 with steps of 0.025 and log$_{10}$ R from -2 to -1 with steps of 0.05 as shown on Figure \ref{fig : new_grid} where the paths of the Sun at different ages are represented, together with the locations of the  base of the convective zone. Moreover the Z grids also have been increased to better adapt to the present solar composition, Z= 0.015 has been added and some interpolations for 0.013 and 0.017. 

\begin{figure*}
\hspace{-1cm}\includegraphics[scale=0.50]{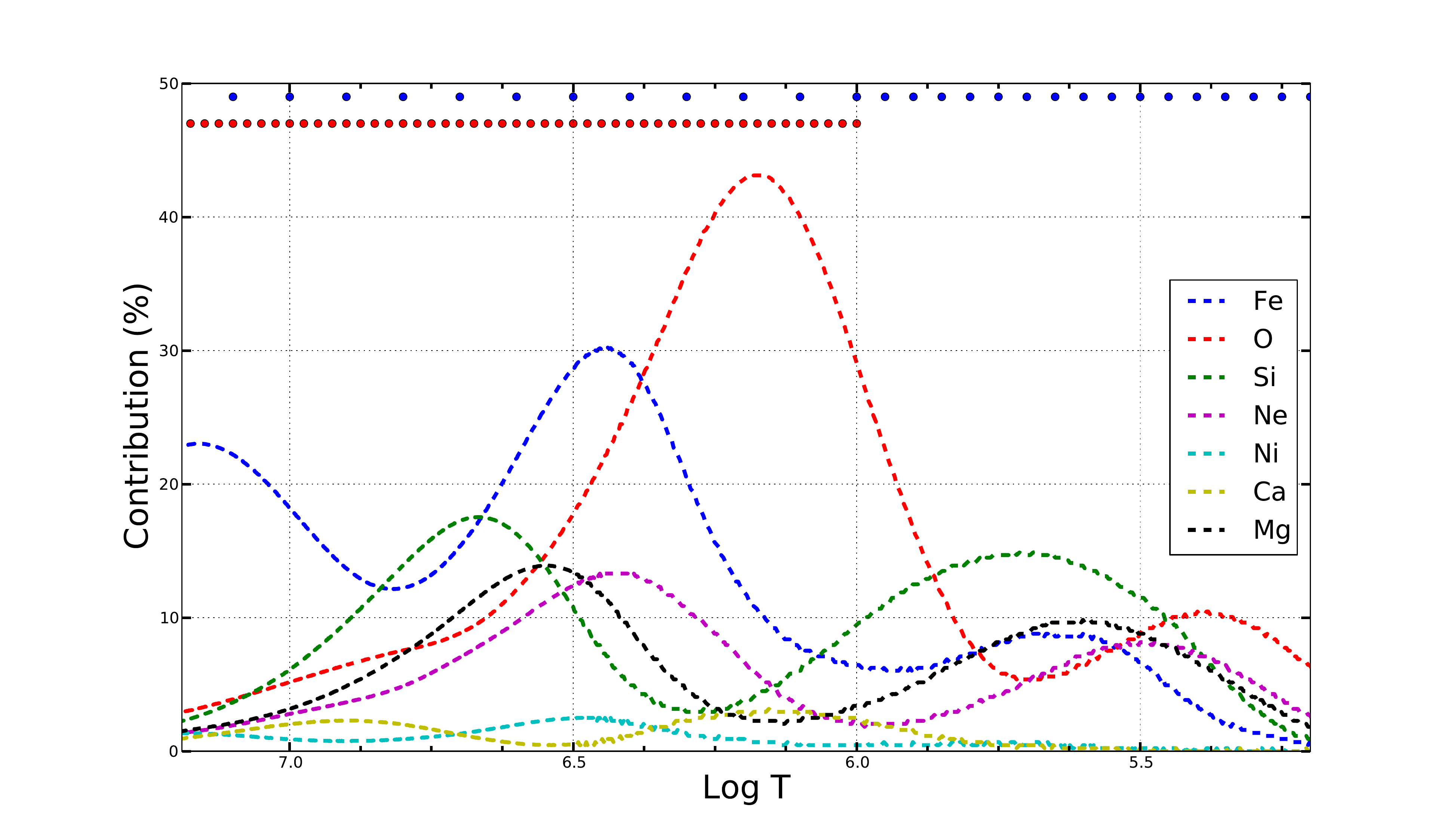} 
 \caption{Relative contribution of the most important heavy elements to the total Rosseland mean opacity (including H and He) for the internal conditions of the present Sun, the composition from Asplund et al. (2009), using OP opacities. OPAS (red circles) and OPAL (blue circles) grids in temperature are indicated in the upper part of the diagram.  }
   \label{fig : contrib_sol}
\end{figure*}

Figure \ref{fig : contrib_sol} recalls the contributions of the most important heavy elements to the global opacity (including H and He). This figure has been realised with OP opacity calculations as monochromatic calculations are available for the different elements. The temperature grids corresponding to OPAL (blue circles) and OPAS (red circles), are also indicated. As one can see, each elementary contribution has a specific shape but a spline  interpolation through the OPAL (or OP) tables 
with only 7-8 points in temperature in the whole radiative zone of the Sun could produce some smoothing effect which do not allow to explore the whole potentiality of the seismic results. It is why the OPAS tables have been designed to significantly improve the interpolation procedure for trying to extract some inner composition signatures from seismology (mainly presently for the Sun) as it was mentioned as an objective before the launch of SoHO \citep{Turck1992}. 

Indeed the uncertainty on the sound speed is about 10$^{-4}$ while its radial location uncertainty varies from 1.5 to 3 \% in radius from the BCZ to the center. Hence a small number of  opacity points doesn't seem sufficient to precisely probe the composition of this region since the Rosseland mean values are significantly dependent on the ionisation state of each element \citep[see][]{Turck1993}.
The OPAL Rosseland mean opacity varies  between two consecutive points of the grid by about 25\% with a change of more than a factor 10 between the center to the BCZ of a solar model. The fine mesh of OPAS presents only a 6\% of Rosseland mean opacity variation between two consecutive points of the grids so the interpolation between points (when introduced in the computation of solar models) will be more accurate. As a consequence, OPAS tables shall present the potential to interpret with a better sensitivity changes of slope in the sound speed profile due to the different opacity processes (recalled in section 2) coming from different element contributions. These tables shall also have the potential to develop inversion of composition inside the radiative zone as it has been possible for the equation of state in the subsurface layers of the Sun \citep{BasuChristensen}. 

Before, one needs first to see how the absolute differences between OPAS and most commonly adopted tables act on the solar model. This is why in this letter we compare the structures of solar models computed with OPAS, OPAL and OP tables  by using the same opacity mesh in each case, i.e. adopting the OPAL standard one (see beginning of this section). Doing so, we do not introduce any adding effect of interpolation that could be difficult to dissociate from physical processes.  We use in that aim two evolution codes popular in the asteroseismic community. 

From Mondet et al. 2015, we know that the OPAS calculations do not differ by more than 10\% from the OPAL ones. In the present study, no more than 6\% differences are observed between OPAS and OPAL calculations for solar conditions so one needs to be cautious in the conclusions we get.

\section{New solar models using the OPAS tables}
In this section we compare SSM computed for the most recent composition \citep{Asplund2009} with two different stellar evolution codes MESA and CLES.  We compare first the impact between the use of OPAS or OPAL tables, as OPAL tables are considered as the best effort in opacities done for solar and solar-type stellar applications. The use of two codes guarantees that the observed effects are really due to the new physics taken into account in the opacity calculations. This precaution is necessary as the differences between the two tables are not so large.

\subsection{The MESA characteristics}
MESA (Modules for Experiments in Stellar Astrophysics) \citep{Paxton2011, Paxton2013,Paxton2015} is a recent stellar evolution code performed for extensive use in the HR diagram. This code is now largely used in the asteroseismic community due to its reliability, its extensive access to a large range of mass and  evolution stage. The rapid progress in the introduction of the physical inputs due to its international use lets it very attractive for a lot of astrophysical applications. 

In the present study we use version 4906 of the code and adopt the following physics input: the MLT theory \citep{Bohm1958}, the OPAL EOS \citep{RogersNayfonov2002}, 
OPAL opacity tables extended to low T and $\rho$ \citep{Ferguson2005}. Nuclear reactions are taken from NACRE \citep{Angulo1999} and 
the microscopic diffusion of all the elements uses the subroutine of  Thoul et al. (1994). The MESA atmosphere model \citep{Paxton2011} comes from tables performed by Castelli \& Kurucz (2003), using the solar composition of Grevesse and Noels (1993).


\subsection{The CLES characteristics}
The stellar evolution code CLES (Code Li\'egeois d'Evolution Stellaire \citep{Scuflaire2008}) has been developed mainly mainly for main sequence  studies and seismic interpretation, and for instance has been compared in detail with the CESAM code \citep[see detailed comparisons in][]{Montalban2008}. An additional smoothing of the opacity tables before their use in the evolution code is an option in CLES. As we observe that it can artificially reduce the values of the opacity, we do not include such treatment in the present study. We use the same physics input than for the MESA computations except that the treatment of the microscopic diffusion only considers three elements: H, He, Fe (all elements heavier than He are treated as Fe). The code uses interpolation in models of
atmosphere  \citep[see][]{Kurucz1998} and performs a smooth junction between interior and atmosphere at T=T$_{eff}$
of the model, with the same limitation than MESA.


\subsection{Use of OPAS tables in the stellar evolution codes}
We have built several calibrated solar models with CLES and MESA  using OPAL and OPAS tables. In this second case and since OPAS tables extend over a limited range of log$_{10}$ T and log$_{10}$ R values, at each mesh point of OPAL tables where there is an existing OPAS calculation, the OPAL opacity value is replaced by the corresponding OPAS value. The OPAL values are adopted for points outside the OPAS domain, but we note that there is no transition in tables due to the mixing of OPAL and OPAS information as the OPAS tables cover the whole solar radiative zone study.

Table 1 summarises the quantities of interest for the calibrated solar models we have computed.
One notices  that, in both cases, the base of the convective zone becomes closer
to the seismic results (0.713 $\pm$ .001 R$_\odot$, \cite{Basu1997}) with the OPAS tables. The initial helium abundance also decreases when using OPAS tables in both cases.
\begin{table}
\begin{center}
\caption{Comparison between solar MESA and CLES models including OPAL, OPAS or OP opacity calculations and the most recent composition \citep{Asplund2009} . Y$_{0}$ is the initial helium , $\alpha$ the MLT value,  $Z/X_{S}$ the surface metallic/ hydrogen ratio at the present age, R$_{CZ}$ the position of the base of the convective zone and T$_{C}$ the central temperature.}
\vspace{5mm}
\begin{tabular}{|c||c|c|c|c|c|}
\hline
  & MESA-OPAL & MESA-OPAS & CLES-OPAL & CLES-OPAS & CLES-OP \\
\hline
\hline
Y$_0$ & 0.2654 & 0.2611 & 0.2681 & 0.2636 & 0.2666\\
\hline
$\alpha$ & 1.77 & 1.79 & 1.75 & 1.76  & 1.76\\ 
\hline
(Z/X)$_{S}$ & 0.01816 & 0.01815& 0.01810 & 0.01810 & 0.01810 \\ 
\hline
\hline
R$_{CZ} \, (R_\odot)$ & 0.729 & 0.723 & 0.724 & 0.719 & 0.723\\
\hline
\hline
T$_{C}$ \, (K) &  15.55 10$^6$ &  15.54 10$^6$& 15.55 10$^6$ & 15.54 10$^6$& 15.52 10$^6$ \\
\hline
\hline
\end{tabular}
\end{center}
\end{table}
With CLES, we have also compared the new results to a solar model using OP tables as was already done for a different solar composition \citep{Scuflaire2008}. 
We note the same tendencies between OP and OPAS than between OPAL and OPAS for the position of the base of the convective zone and for the initial helium.

\section{Sound speed and density profiles compared to helioseismic results}
We have extracted the solar sound speed and density from the previous models and we compare them to the seismic observations  \citep[see all the numbers in][]{TurckLopes}. 

\subsection{Comparison between models using OPAL and OPAS opacities}
Figure 3 shows a clear reduction of the difference between the SSM squared sound speed or density profile and the observed seismic values along one third of the radiative zone below the base of the convective zone,  when one uses the OPAS values in the OPAL tables. The same effect is observed for the two evolutionary codes and can be directly attributed to the change of opacities. 

The observed improvement could be attributed to more complete opacity calculations of iron, nickel and several other low abundant element (with high atomic number) contributors to the Rosseland mean OPAS values. Indeed, near the base of the convection zone, bound-bound processes are important for these elements.  Even if it is difficult to conclude without a detailed comparison of the spectra, it is important to recall that 6 \% on the mean value could come from 30-40\% differences on some specific elements, \citep[see][]{Blancard2012}. Moreover the difference in absolute values of the position of the BCZ could originate from the way the opacities are used (smoothing or not smoothing of the opacities) in the two codes and on the difference in the treatment of the microscopic diffusion. This point will be studied in details in a more complete paper.

On the contrary, in the nuclear region and slightly above it, the agreement is slightly worse and the central temperature slightly reduced as shown in table 1 due to a reduction by less than 5\% of the Rosseland mean values of OPAS compared to OPAL ones  \citep[as shown on Figure 5 by][]{Mondet2015}. The reasons have not been studied in details but a check of the reliability of these calculations would be useful. Some experimental validation to study the plasma effects has been already studied \citep{LePennec2015}.


\begin{figure*}
\hspace{-2cm} \includegraphics[scale=0.50]{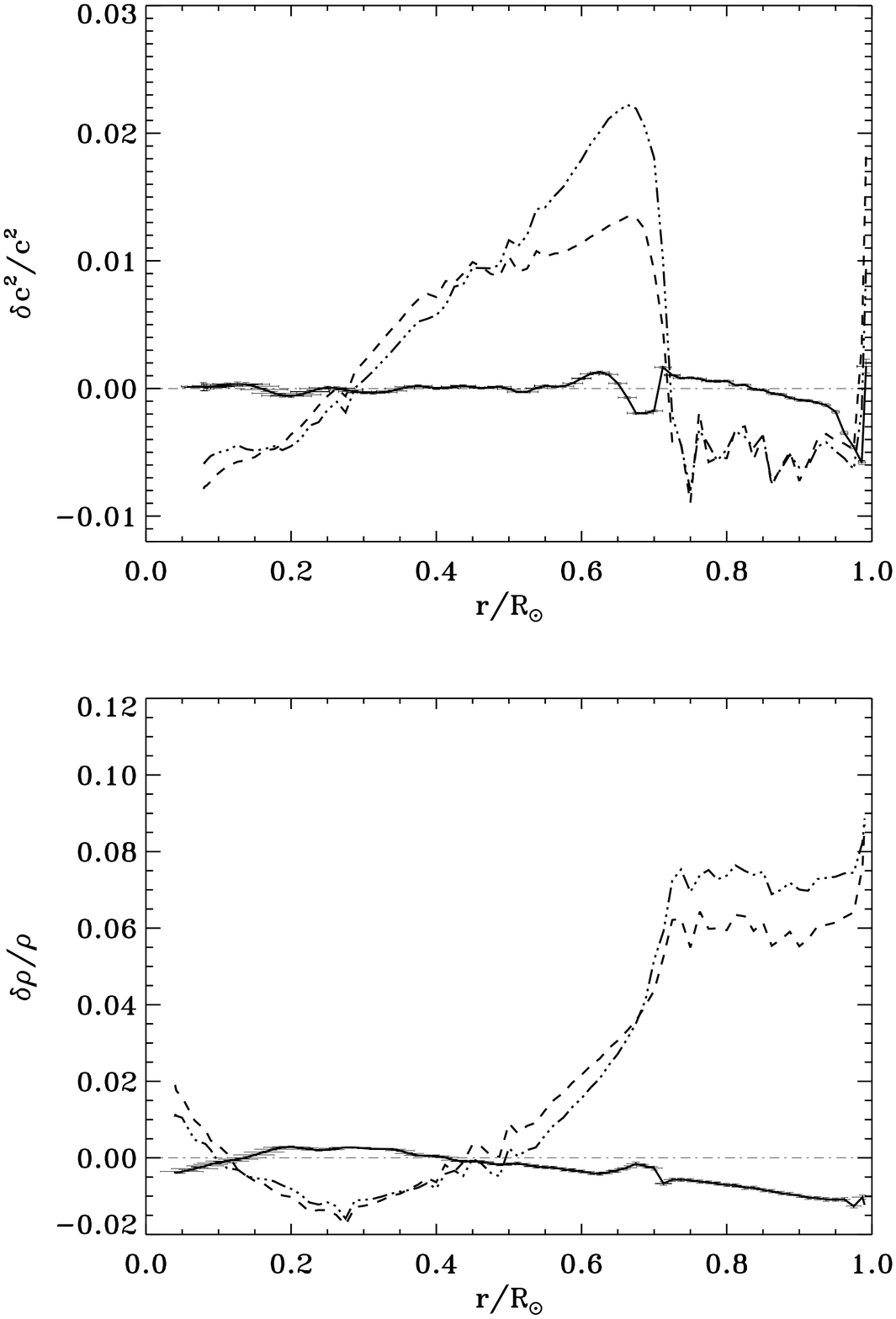}
\hspace{-0.5cm} \includegraphics[scale=0.52]{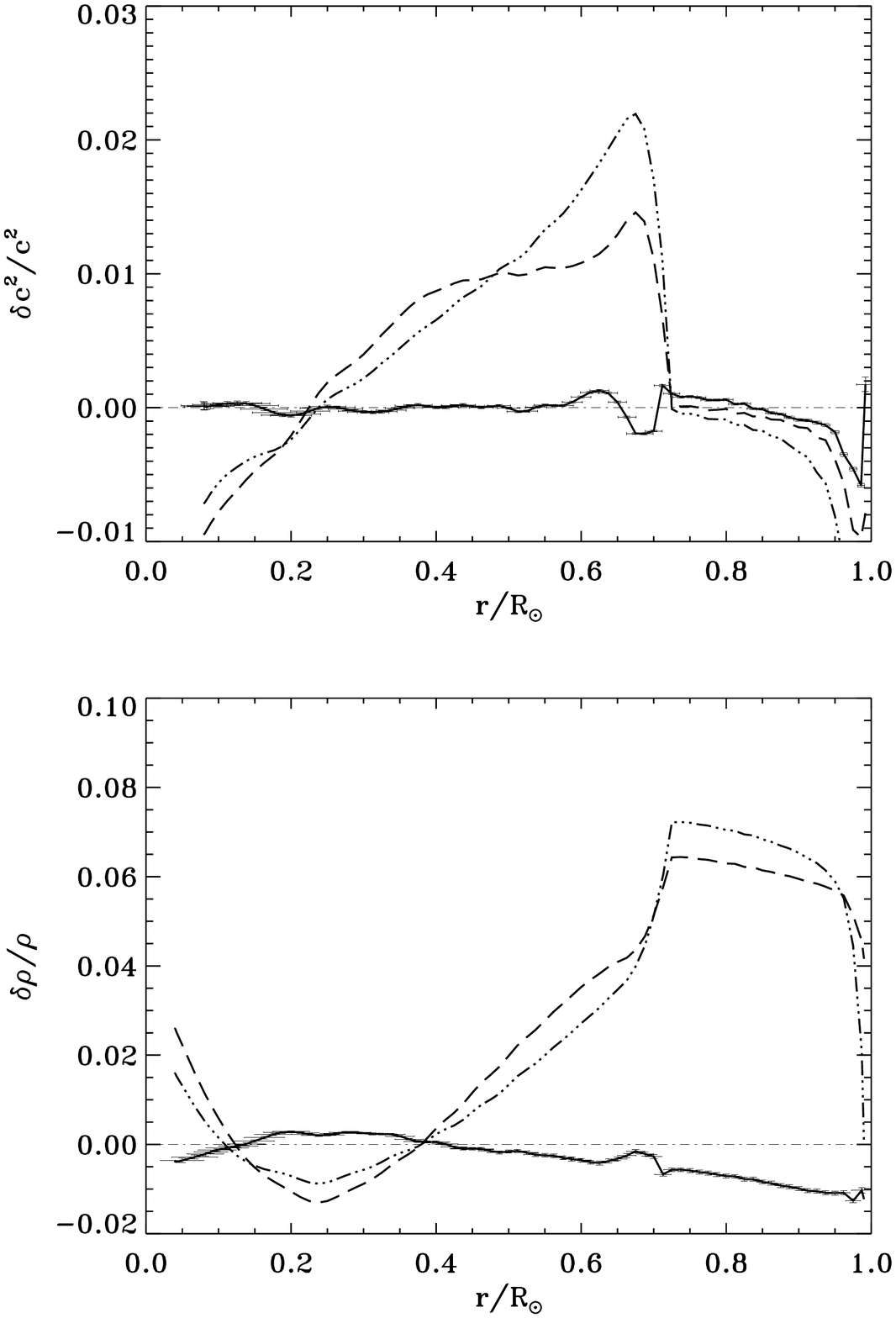}
\caption{Left: Difference between the observed squared sound speed and density profiles with those obtained with a SSM model of MESA using OPAL ($-... -$ line) or OPAS ({$---$} line). Right: Idem for a SSM model performed with CLES. 
The full line corresponds to a seismic model, the associated error bars are extracted from the inversion done using GOLF+MDI aboard SOHO observations \citep[see][for numbers and details]{TurckLopes}.}
\end{figure*}

\subsection{Comparison between models using OP and OPAS opacities }
One notes on Figure 4 for models computed with CLES that the improvements, passing from OP to OPAS tables, seem really smaller. Nevertheless, Table 1 shows the same progress for the position of the base of the convective zone. 
\begin{figure*}
 \begin{center}
\hspace{-0.5cm} \includegraphics[scale=0.52]{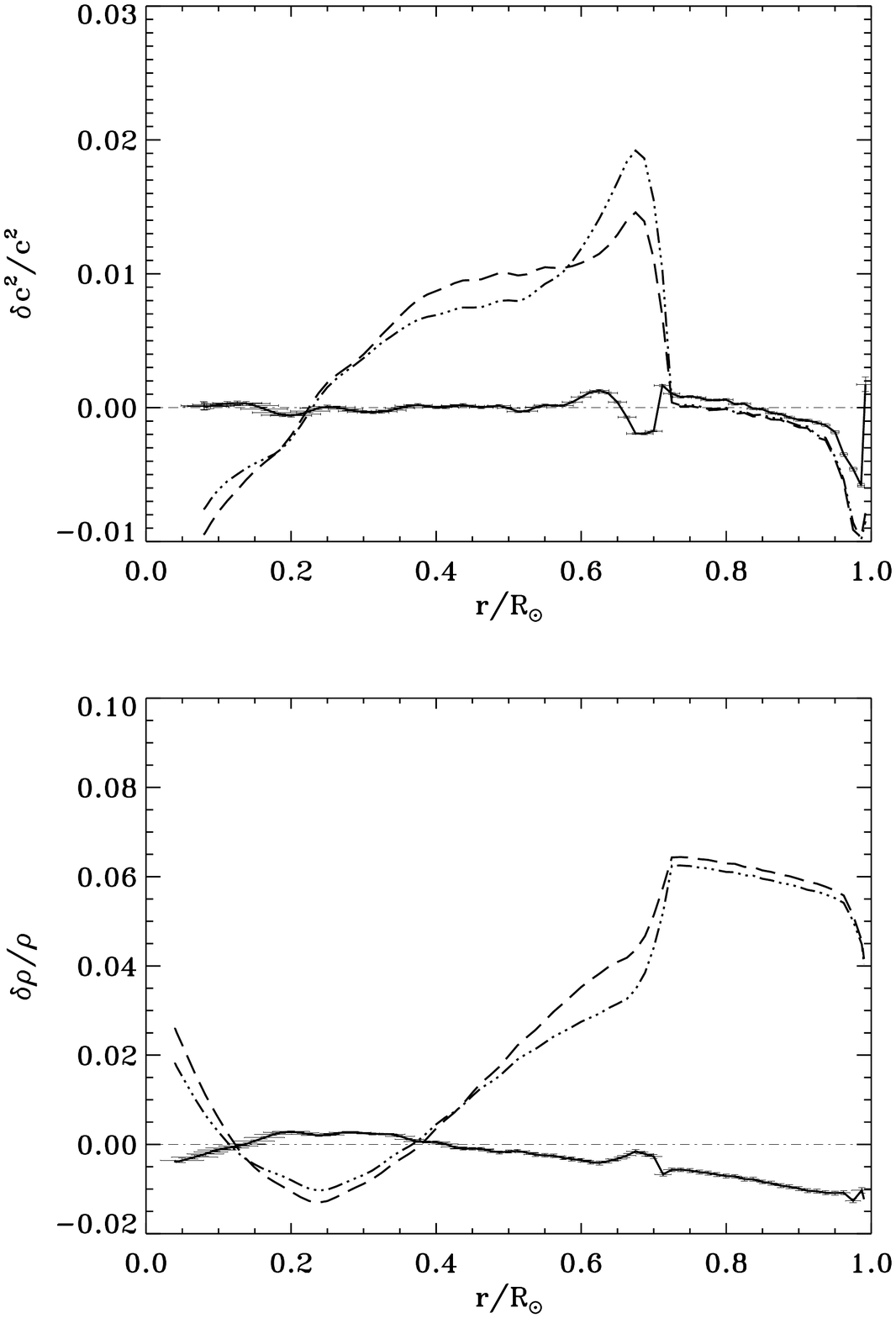}
\caption{Difference between the observed squared sound speed and density profiles \citep{TurckLopes} with those obtained with a SSM model of CLES and using OP opacity ($-... -$ line) or OPAS ($---$ line). Same comments than in Figure 3.}
 \end{center}
\end{figure*}
In fact OPAS monochromatic opacity calculations differ from OP calculations in the  description of the Stark profile of the He-alpha line \citep{Blancard2012}.
The width is greater in OP calculations and this effect increases with Z.
Indeed oxygen, neon, magnesium and silicium are affected by this effect with a resulting larger opacities for these elements in the case of OP calculations. 
On the contrary, in the case of iron, due to the greater number of considered excited states, OPAS calculations are greater than OP ones.
So, as the differences in oxygen and iron opacities are in opposite sign, the recent progress performed by the new generation of opacity codes is not clearly visible but the surprising result on the Z pinch experiment does not favor the OP opacity calculations on iron compared to OPAS ones \citep{Bailey2015}.

 
\section{Conclusion and perspectives}
New refined opacity tables are now available for the modelling of the Sun and solar-like stars \citep{Mondet2015}. In this paper we show the physical change obtained in using OPAS tables in OPAL or OP grids with the same mesh. These improved calculations present opacity differences with OPAL of no more than $\pm$ 5-6 \% in the conditions used in the present study. Such changes already reduce the differences with the seismic observations when compared to the use of OPAL tables, both for the base of the convective zone and for the radiative sound speed profile in the radiative region, it could be attributed to a more complete treatment of the bound-bound processes of the iron group elements. The progress in comparison with OP is also shown but it is largely reduced due a compensation effect between iron and oxygen. Nevertheless OP is not preferred to OPAS when the Z machine recent experiment is taken into account. 


The present study shows the direct effect of improvement in the opacity calculations for some elements of the iron group. The interest of the OPAS tables goes beyond the present study as the fine grids in both log$_{10}$ T, log$_{10}$ R and Z will improve the interpolation through the tables for Sun and solar-like stars. The fine meshes of OPAS will be used to try to extract some specific signatures of the deep composition of the Sun. This work is in progress and a more complete study using the potential of the fine meshes of OPAS will be discussed in a more detailed paper (Salmon et al. in preparation). The present results strongly encourage complementary experimental studies on high energy density laser facilities both on iron and oxygen \citep{Keiter2013, LePennec2015}.

\begin{acknowledgments}
This work has been done in the framework of the french ANR OPACITY. We would like to thank also J. Montalb\'an for her great expertise in the use of the CLES code. We thank also the referees for their judicious remarks which lead to an improved letter.
\end{acknowledgments}

\bibliographystyle{plainnat}

\end{document}